\documentclass{elsart}
\usepackage{graphicx}

\newcommand{\Jv}{\mbox{\boldmath$J$}}
\newcommand{\xiv}{\mbox{\boldmath$\xi$}}
\newcommand{\sigv}{\mbox{\boldmath$\sigma$}}

\begin{document}

\begin{frontmatter}

\title{Complex and Non-Complex Phase Structures
in Models of Spin Glasses\\
and Information Processing}
\author{Hidetoshi Nishimori}
\address{Department of Physics, Tokyo Institute of Technology,
Oh-okayama, Meguro-ku, Tokyo 152-8551, Japan}

\begin{abstract}
The gauge theory of spin glasses and
statistical-mechanical formulation of error-correcting codes
are reviewed with an emphasis on their similarities.
For the gauge theory, we explain the functional identities on dynamical
autocorrelation functions and on the distribution functions of
order parameters.
These functional identities restrict the possibilities
of slow dynamics and complex structure of the phase space.
An inequality for error-correcting codes is shown
to be interpreted to indicate non-monotonicity of spin orientation
as a function of the temperature in spin glasses.

\end{abstract}


\end{frontmatter}

%%%%%%%%%%%%%%%%%%%%%%%%%%%%
\section{Introduction}

The mean-field theory of spin glasses is well established
\cite{Nishimori_OUP}.
There exists a low-temperature spin glass phase
with complex structure of the phase space with such characteristics as
replica symmetry breaking, many valleys and slow dynamics.
In contrast, investigations of realistic finite-dimensional systems
have been hampered by difficulties of analytic treatment,
and numerical methods have been the predominant tool of research.
An interesting exception is the gauge theory which exploits
gauge symmetry of the system to derive many exact/rigorous
results such as the exact expression of the internal energy,
a bound on the specific heat and various identities
for correlation functions \cite{Nishimori_OUP,Nishimori81}.

It may be a little bit surprising that this gauge theory is
closely related with some problems of information processing
such as error-correcting codes and image restoration.
Suppose that one transmits digital information
(a bit sequence) through a noisy channel.
An important goal of error-correcting codes is to devise
an efficient way to encode and decode the information in order
to remove as much noise as possible from the output of the channel.
This problem can be rewritten as an Ising spin glass \cite{Sourlas89}:
A bit sequence is naturally expressed as an Ising spin configuration,
and noise corresponds to randomness in exchange interactions.
Then the statistical inference of the original information,
given the noisy output of the channel, is found to be equivalent
to the statistical mechanics of an Ising spin glass.

A similar analogy exists between image restoration, in which
one tries to remove noise from a given digital image,
and the statistical mechanics of a spin system in random fields.
It is the purpose of this contribution to review the gauge
theory of spin glasses and information processing problems,
error-correcting codes in particular, with
emphasis on the similarities between these superficially quite different
fields \cite{Nishimori_OUP}.

The gauge theory of spin glasses and error-correcting codes
are reviewed in sections 2 and 3, respectively,
with a few new results and viewpoints.
Summary is given in section 4.

%%%%%%%%%%%%%%%%%%%%%%%%%%%%%%%%%
\section{Gauge theory of spin glasses}
%%%%%%%%%%%%%%%%%%%%%%%%%%%%%%%%%
The gauge theory of spin glasses is a powerful theoretical framework
to derive a number of exact/rigorous results on models
of spin glasses using gauge symmetry of the system.
It does not answer directly the question of the existence
or absence of the spin glass phase in finite dimensions.
Nevertheless this theory leads to various results which set
strong constraints on the possibilities of the existence
region of the spin glass phase in the phase diagram and
other non-trivial conclusions
\cite{Nishimori_OUP,Nishimori81,Ozeki-Nishimori93}.

%%%%%%%%%%%%%%%%%%%%%%%%%%
\subsection{System and its symmetry}
%%%%%%%%%%%%%%%%%%%%%%%%%%

Let us consider the $\pm J$ Ising model with the Hamiltonian
  \begin{equation}
     H=-\sum_{\langle ij \rangle} J_{ij}S_i S_j,
      \label{Ising_Hamiltonian}
  \end{equation}
where $J_{ij}$, a quenched random variable, is either $J$
(with probability $p$) or $-J$ (with probability $1-p$).
The sum over $\langle ij \rangle$ runs over appropriate pairs of
sites, such as (but not restricted to)
nearest neighbours on a regular lattice.
This probability distribution can be written in a compact form as
 \begin{equation}
    P(J_{ij})=\frac{e^{K_p \tau_{ij}}}{2\cosh K_p},
      \label{Prob_distribution}
 \end{equation}
where $\tau_{ij}(=\pm 1)$ is the sign of $J_{ij}$.
It is easy to check that (\ref{Prob_distribution}) is equal to $p$
if $\tau_{ij}=1$ and is $1-p$ when $\tau_{ij}=-1$ if we define the parameter
$K_p$ by $e^{-2K_p}=(1-p)/p$.

The Hamiltonian (\ref{Ising_Hamiltonian}) is invariant under
the gauge transformation
  \begin{equation}
    S_i \to S_i\sigma_i,~~J_{ij}\to J_{ij}\sigma_i \sigma_j,
     \label{g-transformation}
  \end{equation}
where $\{ \sigma_i\}$ is a set of Ising variables fixed arbitrarily
at each site.

%%%%%%%%%%%%%%%%%%%%%%%%%%
\subsection{Exact energy}
%%%%%%%%%%%%%%%%%%%%%%%%%%

\begin{figure}
  %\vspace*{4cm}
  \begin{center}
   \includegraphics[width=.35\linewidth]{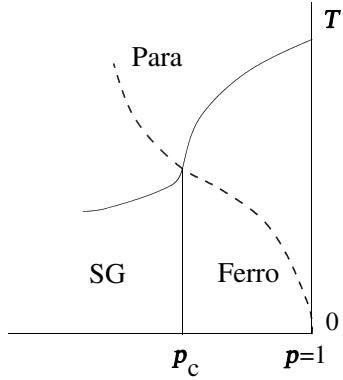}
  \end{center}
  \caption{Phase diagram of the $\pm J$ model and the
   Nishimori line (dashed).}
  \label{fig:phase_diagram1}
\end{figure}

It is possible to derive the exact expression of the internal energy,
averaged over the probability (\ref{Prob_distribution}), under a
simple condition between the temperature and the probability $p$.
The definition of the internal energy is
  \begin{equation}
   [E]=\sum_{\{ \tau_{ij}\}} \prod_{\langle ij \rangle}
    P(J_{ij}) \cdot \frac{\sum_{\{ S_i\}} H e^{-\beta H}}
                       {\sum_{\{ S_i\}} e^{-\beta H}},
            \label{E_def}
   \end{equation}
where the square brackets $[\cdots ]$ denote the configurational average.
The gauge transformation (\ref{g-transformation}) just redefines
the running variables $\{ \tau_{ij}\}$ and $\{ S_i\}$ and
therefore does not affect the value of the internal energy.
Thus (\ref{E_def}) is rewritten as
  \begin{equation}
    [E]= \sum_{\tau} \frac{e^{K_p \sum \tau_{ij}\sigma_i \sigma_j}}
   {(2\cosh K_p)^{N_B}}\cdot
         \frac{\sum_S H e^{K\sum \tau_{ij}S_i S_j}}
              {\sum_S   e^{K\sum \tau_{ij}S_i S_j}},
      \label{E_gauge}
  \end{equation}
where $N_B$ is the number of bonds in the system and $K=\beta J$.
Since the value of (\ref{E_gauge}) is independent of the choice of
$\{\sigma_i\}$, we may sum the right hand side
of (\ref{E_gauge}) over all possible values
of $\{\sigma_i\}$ and divide the result by $2^N$ without changing
the value of $[E]$:
  \begin{equation}
     [E]=\frac{1}{(2\cosh K_p)^{N_B} 2^N} \sum_{\tau} \sum_{\sigma}
  e^{K_p\sum \tau_{ij} \sigma_i \sigma_j}
   \cdot
         \frac{\sum_S H e^{K\sum \tau_{ij}S_i S_j}}
              {\sum_S   e^{K\sum \tau_{ij}S_i S_j}}.
  \end{equation}
It is clear here that the denominator on the right hand side
cancels with the sum over $\sigma$ if $K=K_p$, which makes it
possible to evaluate the energy explicitly:
  \begin{eqnarray}
    [E]&=&\frac{1}{(2\cosh K_p)^{N_B} 2^N} \sum_{\tau} \sum_{S}
       H e^{K\sum \tau_{ij}S_i S_j}
          \nonumber\\
    &=& \frac{1}{(2\cosh K_p)^{N_B} 2^N} 
       \left(-\frac{\partial}{\partial \beta}\right)
     \sum_{\tau} \sum_{S} e^{K\sum \tau_{ij}S_i S_j}
       \nonumber\\
    &=&\frac{1}{(2\cosh K_p)^{N_B} 2^N} 
       \left(-\frac{\partial}{\partial \beta}\right)
        \sum_{S} (2\cosh K)^{N_B}
        \nonumber\\
    &=& -N_B J \tanh K.
          \label{exact_energy}
   \end{eqnarray}
The only condition for the above manipulations to be valid is the
relation $K=K_p$ to relate the temperature with the probability,
defining a line in the phase diagram, the Nishimori line shown
dashed in Fig. \ref{fig:phase_diagram1}.
%------------------
%------------------
Note that the lattice structure, range of interaction, or
the spatial dimension are all arbitrary.
So the present exact solution (\ref{exact_energy}) is very generic.

Similar arguments lead to an upper bound on the specific heat
and various identities for static correlation functions under the
condition $K=K_p$.

%%%%%%%%%%%%%%%%%%%%%%%%%%%
\subsection{Absence of slow dynamics}

\begin{figure}
  %\vspace*{4cm}
  \begin{center}
   \includegraphics[width=.5\linewidth]{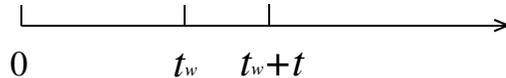}
  \end{center}
  \caption{Two measurement times of the dynamical correlation function
       on the right hand side of (\ref{dynamical_correlation}).}
  \label{fig:dynamical_corr}
\end{figure}

An interesting relation can be established for dynamical
correlation functions using the gauge theory \cite{Ozeki95,Ozeki97}:
  \begin{equation}
    \left[ \left\langle S_i(t_w) S_i(t+t_w) \right\rangle_{K_p}^{\rm F}
  \right]
  = \left[ \left\langle S_i(0) S_i(t) \right\rangle_{K_p}^{\rm eq} \right].
  \label{dynamical_correlation}
 \end{equation}
The right hand side is the equilibrium autocorrelation function
at temperature $K=K_p$.
The left hand side is a non-equilibrium
correlation function starting from the completely ferromagnetic state
at time 0.
The first measurement is performed after waiting time $t_w$  and the second
at $t_w+t$ (Fig. \ref{fig:dynamical_corr}).
%------------------
%------------------
%
If the systems shows anomalously slow dynamics, the non-equilibrium
correlation function, the left hand side, will depend on the
measurement starting time (the waiting time $t_w$)
as it actually happens in the spin glass phase of the mean-field model.
Otherwise, the system relaxes quickly to equilibrium, implying
time-translation invariance, or equivalently, independence of the
left hand side of $t_w$.
The identity (\ref{dynamical_correlation}) proves that the
latter possibility is the case since the right hand side does not
depend on $t_w$.
Therefore, any spin glass system does not show slow
dynamics on the line $K=K_p$.
The proof of (\ref{dynamical_correlation}) requires some background of
the kinetic Ising model and we refer the reader to the original papers.

Note that the above argument is not a proof of the absence of
a spin glass phase in general.
What has been shown is that a phase with slow dynamics
should exist, if it does at all, away from the Nishimori line.

%%%%%%%%%%%%%%%%%%%%%%%%%
\subsection{Distribution functions of order parameters}
%%%%%%%%%%%%%%%%%%%%%%%%%
An interesting functional
identity $P_m(x)=P_q(x)$ can be proved under the same condition
$K=K_p$ \cite{Nishimori-Sherrington}.
Here, $P_m(x)$ is the distribution function of magnetization and
$P_q(x)$ is the distribution of spin glass order parameter,
\begin{equation}
 P_m(x)=\left[ \frac{\sum_{S}\delta \left(x-\frac{1}{N}\sum_i S_i \right)
  e^{-\beta H}}{\sum_{S}e^{-\beta H}}\right]
  \label{Ising-Pm}
\end{equation}
and 
\begin{equation}
 P_q(x)=\left[ \frac{\sum_{S}\sum_{\sigma}
    \delta \left(x-\frac{1}{N}\sum_i S_i \sigma_i \right)
  e^{-\beta H(S)-\beta H(\sigma )}}
    {\sum_{S}\sum_{\sigma}e^{-\beta H(S)-\beta H(\sigma )}}
   \right].
   \label{Ising-Pq}
\end{equation}
The proof of the identity $P_m(x)=P_q(x)$ is relatively straightforward if we
apply the gauge transformation to the expression (\ref{Ising-Pm}) and
use the condition $K=K_p$.
After the gauge transformation, (\ref{Ising-Pm}) is expressed as
  \begin{eqnarray}
     P_m(x)&=& \frac{1}{(2\cosh K_p)^{N_B} 2^N} \sum_{\tau} \sum_{\sigma}
      e^{K_p \sum \tau_{ij}\sigma_i \sigma_j} \nonumber\\
     &\cdot & \frac{\sum_{\sigma}\sum_S 
        \delta \left(x-\frac{1}{N}\sum_i S_i \sigma_i \right)
       e^{K_p \sum \tau_{ij}\sigma_i \sigma_j}
       e^{K   \sum \tau_{ij}S_i S_j}}
      {\sum_{\sigma}e^{K_p \sum \tau_{ij}\sigma_i \sigma_j}
       \sum_S
       e^{K   \sum \tau_{ij}S_i S_j}}.
    \end{eqnarray}
The last expression is equivalent to
(\ref{Ising-Pq}) if $K=K_p$.

Since the distribution function of magnetization $P_m(x)$ is always
composed of at most two simple delta functions located at $\pm m$,
the identity $P_m(x)=P_q(x)$ implies that the distribution of
spin glass order parameter $P_q(x)$ also has the same simple structure.
This excludes the possibility that a complex structure of the phase space,
which should be reflected in a non-trivial functional form
of $P_q(x)$, exists on the line $K=K_p$.
Thus a spin glass (or mixed ferromagnetic) phase of the mean-field type,
if it exists, should lie away from
the line in the phase diagram.
This observation reinforces the conclusion of the previous
subsection on dynamics that there is no mean-field type spin glass phase
on the Nishimori line.

%%%%%%%%%%%%%%%%%%%%%%%%%%%%
\section{Error-correcting codes}
%%%%%%%%%%%%%%%%%%%%%%%%%%%%

The arguments in the previous section on the gauge theory of
spin glasses are in close formal similarity to some problems
of information processing.
We demonstrate it with error-correcting codes as
a typical example \cite{Nishimori_OUP}.

%%%%%%%%%%%%%%%%%%%%%%%%%%%%
\subsection{Transmission of information}
%%%%%%%%%%%%%%%%%%%%%%%%%%%%

\begin{figure}
  %\vspace*{4cm}
  \begin{center}
   \includegraphics[width=.8\linewidth]{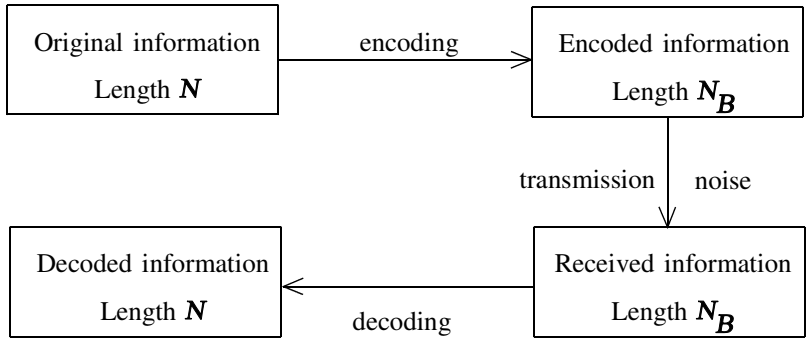}
  \end{center}
  \caption{The original information of $N$ bits is encoded to a binary sequence
   of $N_B (>N)$ bits which is then transmitted through the noisy channel.
   The output of the channel is decoded to infer the original information.}
  \label{fig:ECC}
\end{figure}

The goal of the theory of error-correcting codes is to find efficient
methods to infer the original information (bit sequence) from the
output of a noisy transmission channel.
For this purpose it is known to be necessary to introduce redundancy
before sending the information through the channel,
a process called encoding (or coding), see Fig. \ref{fig:ECC}.
At the receiving end of the channel, one decodes the signal to retrieve
the original information making full use of the redundancy.
Shannon's channel coding theorem sets a lower limit on the
redundancy in order to retrieve the original information without errors,
given the noise strength of the channel.
%------------------
%------------------
%

%%%%%%%%%%%%%%%%%%%%%%%%%%%%%%
\subsection{Ising representation}
%%%%%%%%%%%%%%%%%%%%%%%%%%%%%%

It is convenient to formulate the problem in terms of Ising spins
as natural correspondences exist between a bit and
an Ising spin (0 and 1 corresponds to $S_i=(-1)^0=1$ and $S_i=(-1)^1=-1$,
respectively) and between the mod-two addition and the
multiplication (e.g., 1+1=0 is equivalent to $(-1)^1\times (-1)^1=(-1)^0=1$)
\cite{Sourlas89}.
Suppose that the original information one wishes to send is an
Ising spin configuration $\{\xi_i\}$.
Redundancy can be introduced by generating a set of many-body Mattis-type
interactions \cite{Mattis} from the spin configuration,
$J_{ijk\cdots}^0=\xi_i\xi_j\xi_k\cdots$.
Clearly the number of interactions $N_B$ is larger than the number of
Ising spins $N$, implying redundancy.
A specific choice of the set of sites to form interactions $\{ijk\cdots\}$
corresponds to a specific code.
A search for a good code is one of the main targets of coding theory,
but we do not review this problem here and continue the argument
without specifying the set $\{ijk\cdots\}$.

The encoded information $\Jv^0\equiv \{J_{ijk\cdots}^0\}$ is transmitted
through the noisy channel.
Let us focus our attention on a binary symmetric channel in which
the input $J_{ijk\cdots}^0$ is flipped to the opposite sign
$J_{ijk\cdots}= -J_{ijk\cdots}^0$ with error probability $p$.
The task is to infer (decode) the original information $\{\xi_i\}$
from the noisy output of the channel $\Jv\equiv \{J_{ijk\cdots}\}$.
Note that frustration exists in the set of interactions $\Jv$
although $\Jv^0$ (which was generated by the Mattis rule) is not frustrated.
Finite-temperature statistical mechanics of the Ising spin glass
with interactions $\Jv$ serves as a useful
theoretical platform to decode the message as shown in the following
\cite{Sourlas89,Rujan93,Nishimori93,Sourlas94,Iba}.

%%%%%%%%%%%%%%%%%%%%%%%%%%%%%%%%%%%%
\subsection{Bayesian formulation}
%%%%%%%%%%%%%%%%%%%%%%%%%%%%%%%%%%%%

The first step is to express the noise characteristics using
the conditional probability as
  \begin{equation}
     P(J_{ijk\cdots}|J_{ijk\cdots}^0)=
      P(J_{ijk\cdots}|\xi_i\xi_j\xi_k\cdots)=
         \frac{e^{\beta_p J_{ijk\cdots}\xi_i\xi_j\xi_k\cdots}}
              {2\cosh \beta_p}
     \end{equation}
which is equal to $p$ if $J_{ijk\cdots}=-J_{ijk\cdots}^0$
and $1-p$ otherwise if we define $\beta_p$ as
  \begin{equation}
      e^{2\beta_p}=\frac{1-p}{p}.
   \end{equation}
Note that $\beta_p$ is essentially equivalent to $K_p$
appeared in the previous section, the only difference being
that $p$ and $1-p$ are exchanged.
Assuming a memoryless channel, in which each bit is affected by
noise independently of other bits, the conditional probability
of the whole spin configuration is
  \begin{equation}
   P(\Jv |\xiv)=\prod_{\{ijk\cdots\}}
   P(J_{ijk\cdots}|\xi_i\xi_j\xi_k\cdots)
   =\frac{\exp (\beta_p \sum_{\{ijk\cdots\}}
           J_{ijk\cdots}\xi_i\xi_j\xi_k\cdots)}
              {(2\cosh \beta_p)^{N_B}},
   \label{cond_probability}
     \end{equation}
where $N_B$ is the size of the set $\Jv$.

Equation (\ref{cond_probability}) represents the conditional probability
of the output $\Jv$ given the input $\xiv$.
To infer the original information, we need the conditional
probability of the input $\xiv$ given the output $\Jv$.
The Bayes formula is useful for this purpose:
  \begin{equation}
    P(\sigv|\Jv )=\frac{P(\Jv |\sigv)P(\sigv)}
       {\sum_{\sigma}P(\Jv |\sigv)P(\sigv)},
    \label{Bayes}
  \end{equation}
where $\sigv$ is a set of dynamical variables used to infer $\xiv$.

We need the explicit form of the distribution function of the
original information $P(\sigv)$, called the prior, to use the Bayes formula
(\ref{Bayes}) to infer $\sigv$ given $\Jv$.
It can reasonably assumed that $P(\sigv)$ is a constant, a uniform
prior, because one often compresses information before encoding,
which usually generates a very uniform distribution of 0s and 1s.
Then the $\sigv$-dependence of the right hand side comes only from
$P(\Jv |\sigv)$, and thus we find
  \begin{equation}
   P(\sigv|\Jv )\propto P(\Jv |\sigv) \propto
         \exp (\beta_p \sum_{\{ ijk\cdots\}}J_{ijk\cdots}
          \sigma_i\sigma_j\sigma_k\cdots).
   \label{posterior}
   \end{equation}
The right hand side is the Boltzmann factor of the Ising spin glass
with many body interactions which are quenched since $\Jv$ is given
and fixed.
Therefore the problem has been reduced to the statistical mechanics
of the Ising spin glass with the effective coupling $\beta_p$.

%%%%%%%%%%%%%%%%%%%%%%%%%%%%%%%%%%%%
\subsection{Statistical inference}
%%%%%%%%%%%%%%%%%%%%%%%%%%%%%%%%%%%%

There are two typical methods to infer the original information
(spin configuration) from the conditional probability
(\ref{posterior}) called the posterior.
One is the MAP (maximum a posteriori probability) method in which
one chooses the configuration $\sigv$ that maximizes $P(\sigv|\Jv )$.
This is equivalent to the ground-state search of the Ising spin glass
  \begin{equation}
   H=-\sum_{\{ ijk\cdots\}}J_{ijk\cdots}\sigma_i\sigma_j\sigma_k\cdots
  \end{equation}
according to (\ref{posterior}).
Another is the MPM (maximizer of posterior marginals) method;
one first marginalizes the posterior with respect to $\sigma_i$
  \begin{equation}
     P(\sigma_i|\Jv )=\sum_{\sigma_j (j\ne i)}  P(\sigv|\Jv )
       \label{marginal}
  \end{equation}
and chooses $\sigma_i$ to maximize this marginal probability.
If we write the inferred value as $\hat{\xi}_i$, we have
$\hat{\xi}_i={\rm arg~max}_{\sigma_i} P(\sigma_i|\Jv )$.
Such a process is carried out at each $i$.
The MPM is equivalent to looking at the sign of the local magnetization
$\hat{\xi}_i={\rm sgn}\langle \sigma_i \rangle$, where
the thermal average is taken using the Boltzmann factor (\ref{posterior})
of the Ising spin system with effective coupling $\beta_p$.
The MAP may be regarded as the same method
$\hat{\xi}_i={\rm sgn}\langle \sigma_i \rangle$, the only difference
being that the thermal average is now evaluated at the ground state.
Thus these two methods may be called the ground-state decoding (MAP)
and finite temperature decoding (MPM), respectively.

An important measure of decoding performance is the overlap of
the decoded information (spin configuration) $\{\hat{\xi}_i \}$
and the original true information $\{\xi_i\}$,
  \begin{equation}
      M=\frac{1}{N} \left[ \sum_i \xi_i \hat{\xi}_i \right],
       \label{overlap_def}
   \end{equation}
where the square brackets denote the configurational average with
respect to the distribution of noise (or equivalently, quenched
randomness in interactions).
Perfect decoding result gives $M=1$ whereas a random result yields $M=0$.
A larger $M$ represents a better result.

\begin{figure}
  %\vspace*{4cm}
  \begin{center}
   \includegraphics[width=.5\linewidth]{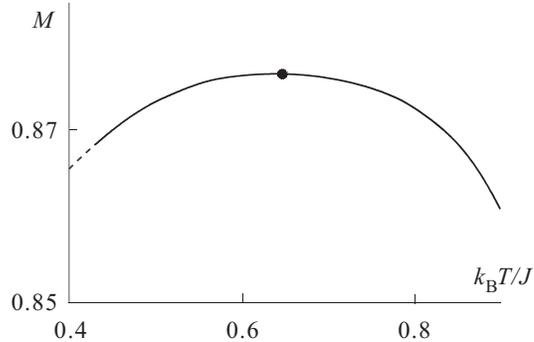}
  \end{center}
  \caption{Overlap as a function of the temperature}
  \label{fig:overlap}
\end{figure}

It is useful to replace the parameter $\beta_p$ in (\ref{posterior})
with a general $\beta$ because, first, the MAP ($\beta
\to \infty$) and the MPM ($\beta =\beta_p$) can be treated within the
same framework and, second, one sometimes does not know the correct
error rate $p$ (and its function $\beta_p$) in applying the MPM.
The overlap is then a function of $\beta$ and $p$.
It turns out that the overlap $M(\beta, p)$ as a function of $\beta$
is not monotonic but reaches the maximum at $\beta_p$
(Fig. \ref{fig:overlap}) \cite{Nishimori93}:
  \begin{equation}
   M(\beta, p) \le M(\beta_p,p).
     \label{M_inequality}
  \end{equation}
%
%------------------
%------------------
This non-monotonicity is not surprising intuitively because the overlap
has been defined as a measure of average bitwise performance
(\ref{overlap_def}) and the MPM has been designed to maximize
the bitwise posterior (\ref{marginal}).
The proof of the inequality (\ref{M_inequality}) will be given
in the next subsection.

%%%%%%%%%%%%%%%%%%%%%%%%%%%%
\subsection{Interpretation as the Ising spin glass}
%%%%%%%%%%%%%%%%%%%%%%%%%%%%

The inequality (\ref{M_inequality}) has the following interesting
interpretation in the context of spin glasses.
Let us consider the problem in terms of the ferromagnetic gauge,
$\xi_i=1~(\forall i)$.
Generality is not lost by this restriction because
the prior has been assumed to be uniform, $P(\sigv)=$const,
and thus all the original configurations have the same
statistical properties, leading to the same average decoding
performance $M(\beta, p)$ for any choice of $\{\xi_i\}$.

In the ferromagnetic gauge the overlap $M(\beta ,p)$ represents the average
of the difference between the numbers of up spins and down spins:
  \begin{equation}
    M(\beta ,p)=\frac{1}{N} \left[ \sum_i \hat{\xi}_i \right]
     =\left[ {\rm sgn}\langle \sigma_i \rangle \right].
       \label{overlap_ferrogauge}
   \end{equation}
It should also be noticed that (\ref{posterior}) is equivalent
to the Boltzmann factor of the $\pm J$ Ising model with
many body interactions only for the
ferromagnetic gauge, strictly speaking.
The reason is that the interaction $J_{ijk\cdots}(=\xi_i\xi_j\xi_k\cdots)$
is equal to $+1$
with probability $1-p$ and to $-1$ with probability $p$
only in the ferromagnetic gauge.
Thus the effective inverse temperature $\beta_p$ 
in (\ref{posterior}) represents the
condition that the system is on the Nishimori line $\beta_p =\beta$
in the ferromagnetic gauge (with $p$ and $1-p$ exchanged
from the notation of section 2).

To prove (\ref{M_inequality}) in the ferromagnetic gauge,
we first write (\ref{overlap_ferrogauge}) explicitly:
 \begin{equation}
   \left[ \frac{\langle \sigma_i \rangle}{|\langle \sigma_i \rangle|}
   \right]
    =\frac{1}{(2\cosh \beta_p)^{N_B}} \sum_{\tau} e^{\beta_p\sum \tau_{ij}}
   \frac{\sum_{\sigma} \sigma_i e^{\beta\sum \tau_{ij} \sigma_i \sigma_j}}
   {|\sum_{\sigma} \sigma_i e^{\beta\sum \tau_{ij} \sigma_i \sigma_j}|}.
  \end{equation}
Then, by applying the gauge transformation (with gauge variable
$\eta_i$) and evaluating the
upper bound by taking the absolute value of the summand, we find
 \begin{eqnarray}
     \left[ \frac{\langle \sigma_i \rangle}{|\langle \sigma_i \rangle|}
   \right]
    &=& \frac{1}{2^N (2\cosh \beta_p)^{N_B}} \sum_{\eta} \sum_{\sigma}
    e^{\beta_p\sum \tau_{ij}\eta_i \eta_j} \langle \eta_i \rangle_{\beta_p}
     \frac{\langle \sigma_i \rangle_{\beta}}
            {|\langle \sigma_i \rangle_{\beta}|}
    \nonumber\\
   &\le& \frac{1}{2^N (2\cosh \beta_p)^{N_B}} \sum_{\eta} \sum_{\sigma}
    e^{\beta_p\sum \tau_{ij}\eta_i \eta_j} |\langle \eta_i \rangle_{\beta_p}|.
  \end{eqnarray}
The last expression is easily seen to be equal to 
$[{\rm sgn}\langle \sigma_i \rangle_{\beta_p}]=
M(\beta_p,p)$ if we apply the gauge transformation to
$[{\rm sgn}\langle \sigma_i \rangle_{\beta_p}]$.

The inequality (\ref{M_inequality}) is interpreted that
the number of up spins, relative to the number of down spins,
is a non-monotonic function of the temperature.
The number of up spins reaches its maximum on the Nishimori
line and then decreases as one further lowers the temperature
(Fig. \ref{fig:overlap}).
In this sense, the spin state on the Nishimori line is
more ordered than the states at any other temperature with the same $p$.

\begin{figure}
  %\vspace*{4cm}
  \begin{center}
   \includegraphics[width=.35\linewidth]{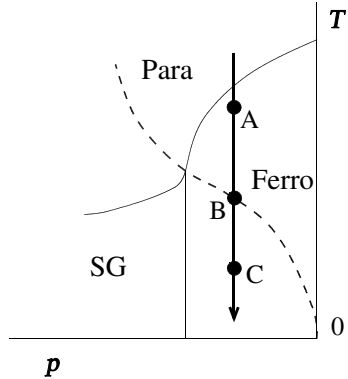}
  \end{center}
  \caption{The spin state is more ordered at point B
   than at A or C.}
  \label{fig:phase_diagram2}
\end{figure}

This is a highly non-trivial result.
As one lowers the temperature from the paramagnetic phase,
the number of up spins becomes to exceed that of down spins
as soon as the system enters the ferromagnetic phase
(point A in Fig. \ref{fig:phase_diagram2})
if one imposes the boundary condition such that the overall
inversion symmetry is broken for the system with two-body
interactions.
\footnote
{
Note that the argument in the present section applies to
the usual case of two-body interactions as well.
}
%------------------
%------------------
The number of up spins reaches the maximum on the
Nishimori line, where, in some sense, thermal fluctuations
are balanced with geometrical frustration (point B).
Then, as one further lowers the temperature, the spin states start
to capture the detailed structure of bond configurations,
which were masked by
thermal fluctuations above the Nishimori line,
and the number of up spins starts to decrease at lowering
the temperatures (point C).

It should be remembered here that the magnetization $m(\beta ,p)
=[\langle \sigma_i \rangle_{\beta}]$, not the
overlap $M(\beta ,p)=[{\rm sgn}\langle \sigma_i \rangle_{\beta}]$,
is of course expected to be a monotonic function
of the temperature.
Only the number of up spins $M(\beta ,p)$,
which ignores the spin-size reduction
due to thermal fluctuations ($|\langle \sigma_i \rangle| <
|{\rm sgn}\langle \sigma_i \rangle |=1$), is non-monotonic.
We may therefore conclude that the Nishimori line marks
the crossover between the purely ferromagnetic region (where
$M$ increases as the temperature is decreased) and
a randomness-dominated region (where $M$ decreases).
This observation is reinforced by the fact that the ferromagnetic
order parameter $m$ is equal to the spin glass order $q$
on the Nishimori line as can be derived from the functional
identity $P_m(x)=P_q(x)$ mentioned in section 2:
We reasonably expect that the ferromagnetic order
dominates $m>q$ above the Nishimori line and the
opposite is true $m<q$ below.

Another interesting and useful aspect concerning the relation
(or equivalence) between the Ising spin glass and error-correcting
codes is the simplicity of phase space and the absence of
slow dynamics on the Nishimori line discussed in section 2.
Practical algorithms of decoding are generally implemented
as iterative solutions of TAP-like equations for local
magnetizations.
Such iterations can be regarded as discrete-time
dynamics of the Ising spin glass.
And the iteration is often carried out on the Nishimori line
to achieve the best performance in the sense of maximum overlap $M$.
Then, if the system shows slow dynamics, the iteration may
not converge to the desired result in a reasonable amount of time.
The results in section 2 guarantee that this does not happen.
\footnote{
There still exists a problem of appropriate choice of the initial condition
in practice \cite{Kabashima}.
}

%%%%%%%%%%%%%%%%%%%%%%%%
\section{Summary}
%%%%%%%%%%%%%%%%%%%%%%%%

Exact/rigorous results have been presented on static and dynamical
properties of the Ising spin glass using the gauge theory.
These results have been shown to restrict the possibilities of complex
structure of the phase space and slow dynamics of the mean-field type.
We have presented a Baysian formulation of error-correcting codes that has
a very close formal similarity to the Ising spin glass.
The best bitwise inference of the original information has been
shown to be achieved by a finite-temperature decoding
corresponding to the Nishimori line derived in the gauge theory
of spin glasses.
We have seen that an inequality on the overlap, which is a measure
of decoding performance of error-correcting codes, has an
interesting interpretation under the context of spin glasses
that the average spin orientation is not a monotonic function of
the temperature.
The spin ordering, in the sense of the number of up spins (ignoring
the spin reduction due to thermal fluctuations), takes its
maximum value not at $T=0$ but at a finite temperature corresponding
to the Nishimori line.
Thus the system of Ising spin glass is most ferromagnetically-ordered not
in the ground state but at this finite temperature even within
the ferromagnetic phase.
%

%%%%%%%%%%%%%%%%%%%%%%%%%%%%

\end{document}